\documentclass[12pt]{article}
\usepackage[table]{xcolor}
\usepackage{colortbl}
\definecolor{lightgray}{gray}{0.9}
\usepackage{graphicx}
\usepackage{lscape}
\usepackage{citesort}
\usepackage{amssymb}
\usepackage{appendix}
\usepackage{multirow}

\usepackage{graphicx}
\usepackage{lscape}
\usepackage{citesort}
\usepackage{amssymb}
\usepackage{appendix}
\usepackage{multirow}

\usepackage{mathrsfs}
\begin{document}
\def\qq{\langle \bar q q \rangle}
\def\uu{\langle \bar u u \rangle}
\def\dd{\langle \bar d d \rangle}
\def\sp{\langle \bar s s \rangle}
\def\GG{\langle g_s^2 G^2 \rangle}
\def\Tr{\mbox{Tr}}
\def\figt#1#2#3{
        \begin{figure}
        $\left. \right.$
        \vspace*{-2cm}
        \begin{center}
        \includegraphics[width=10cm]{#1}
        \end{center}
        \vspace*{-0.2cm}
        \caption{#3}
        \label{#2}
        \end{figure}
    }

\def\figb#1#2#3{
        \begin{figure}
        $\left. \right.$
        \vspace*{-1cm}
        \begin{center}
        \includegraphics[width=10cm]{#1}
        \end{center}
        \vspace*{-0.2cm}
        \caption{#3}
        \label{#2}
        \end{figure}
                }

\def\ds{\displaystyle}
\def\beq{\begin{equation}}
\def\eeq{\end{equation}}
\def\bea{\begin{eqnarray}}
\def\eea{\end{eqnarray}}
\def\beeq{\begin{eqnarray}}
\def\eeeq{\end{eqnarray}}
\def\ve{\vert}
\def\vel{\left|}
\def\ver{\right|}
\def\nnb{\nonumber}
\def\ga{\left(}
\def\dr{\right)}
\def\aga{\left\{}
\def\adr{\right\}}
\def\lla{\left<}
\def\rra{\right>}
\def\rar{\rightarrow}
\def\lrar{\leftrightarrow}
\def\nnb{\nonumber}
\def\la{\langle}
\def\ra{\rangle}
\def\ba{\begin{array}}
\def\ea{\end{array}}
\def\tr{\mbox{Tr}}
\def\ssp{{\Sigma^{*+}}}
\def\sso{{\Sigma^{*0}}}
\def\ssm{{\Sigma^{*-}}}
\def\xis0{{\Xi^{*0}}}
\def\xism{{\Xi^{*-}}}
\def\qs{\la \bar s s \ra}
\def\qu{\la \bar u u \ra}
\def\qd{\la \bar d d \ra}
\def\qq{\la \bar q q \ra}
\def\gGgG{\la g^2 G^2 \ra}
\def\q{\gamma_5 \not\!q}
\def\x{\gamma_5 \not\!x}
\def\g5{\gamma_5}
\def\sb{S_Q^{cf}}
\def\sd{S_d^{be}}
\def\su{S_u^{ad}}
\def\sbp{{S}_Q^{'cf}}
\def\sdp{{S}_d^{'be}}
\def\sup{{S}_u^{'ad}}
\def\ssp{{S}_s^{'??}}

\def\sig{\sigma_{\mu \nu} \gamma_5 p^\mu q^\nu}
\def\fo{f_0(\frac{s_0}{M^2})}
\def\ffi{f_1(\frac{s_0}{M^2})}
\def\fii{f_2(\frac{s_0}{M^2})}
\def\O{{\cal O}}
\def\sl{{\Sigma^0 \Lambda}}
\def\es{\!\!\! &=& \!\!\!}
\def\ap{\!\!\! &\approx& \!\!\!}
\def\md{\!\!\!\! &\mid& \!\!\!\!}
\def\ar{&+& \!\!\!}
\def\ek{&-& \!\!\!}
\def\kek{\!\!\!&-& \!\!\!}
\def\cp{&\times& \!\!\!}
\def\se{\!\!\! &\simeq& \!\!\!}
\def\eqv{&\equiv& \!\!\!}
\def\kpm{&\pm& \!\!\!}
\def\kmp{&\mp& \!\!\!}
\def\mcdot{\!\cdot\!}
\def\erar{&\rightarrow&}
\def\olra{\stackrel{\leftrightarrow}}
\def\ola{\stackrel{\leftarrow}}
\def\ora{\stackrel{\rightarrow}}

\def\simlt{\stackrel{<}{{}_\sim}}
\def\simgt{\stackrel{>}{{}_\sim}}


\title{
         {\Large
                 {\bf
                     The strong $\Lambda_bNB$ and $\Lambda_cND$ vertices 
                 }
         }
      }

\author{\vspace{1cm}\\
{\small K. Azizi$^a$ \thanks {e-mail: kazizi@dogus.edu.tr}\,, Y.
Sarac$^b$
\thanks {e-mail: ysoymak@atilim.edu.tr}\,\,, H.
Sundu$^c$ \thanks {e-mail: hayriye.sundu@kocaeli.edu.tr}} \\
{\small $^a$  Department of Physics, Do\u gu\c s University, Ac{\i}badem-Kad{\i}k\"oy, 34722 Istanbul, Turkey} \\
{\small $^b$ Electrical and Electronics Engineering Department,
Atilim University, 06836 Ankara, Turkey} \\
{\small $^c$ Department of Physics, Kocaeli University, 41380 Izmit,
Turkey}}
\date{}

\begin{titlepage}
\maketitle
\thispagestyle{empty}

\begin{abstract}
We investigate the strong  vertices  among $\Lambda_b$, nucleon and $B$ meson as well as $\Lambda_c$, nucleon and $D$ meson in QCD.  In particular, we calculate the strong coupling constants $g_{\Lambda_bNB}$
 and $g_{\Lambda_cND}$ for different Dirac structures entered the calculations. In the case of $\Lambda_cND$ vertex, the result is compared with the only existing prediction obtained at $Q^2=0$.

\end{abstract}

~~~PACS number(s): 13.30.-a,  13.30.Eg, 11.55.Hx
\end{titlepage}

\section{Introduction}
The last decade   has witnessed to significant experimental
progresses on the spectrum and decay products of the hadrons
containing heavy quarks. These progresses have been stimulated  the
theoretical interests on the spectroscopy of these  baryons  via various methods
 (for some of them see~\cite{Wang a,Wang a1,Wang a2,Wang a3,Grozin,Mathur,Ebert,Karliner,Karliner1,Rosner,kazemazizi} and references therein). For a
better understanding of the heavy flavor physics, it is also necessary to
gain deeper insight into the radiative, strong and weak decays of  the baryons containing a heavy
quark. For some of the related studies, see ~\cite{Diaz,Scholl,Faessler,Patel,An,Aliev,Aliev1,Navarra,Huang,Wang,Azizi,chivili,Cheng,Khodjamirian,Nieves,sarac,Gutsche} and references therein. 

The strong coupling constants are the main ingredients of the strong
interactions of the heavy baryons. To improve our understanding on
the strong interactions among the heavy baryons and
other hadrons and gain knowledge about the nature and structure of the participated 
particles, one needs the accurate determinations of these coupling
constants. In the present study,  we calculate the strong coupling constants $g_{\Lambda_bNB}$
 and $g_{\Lambda_cND}$ within the framework of the  QCD sum rule~\cite{Shifman} as one of the most  powerful and applicable tools to hadron physics. 
These coupling constants are relevant in the bottom and charmed mesons clouds
description of the nucleon which may be used to explain exotic events observed by different 
Collaborations. Besides, in order to exactly determine the modifications in the masses, decay constants and other parameters of the $B$ and $D$ mesons in nuclear medium we should immediately
consider the  contributions of the baryons $\Lambda_{b[c]}$ and $\Sigma_{b[c]}$ in the medium produced by the interactions of $B$ and $D$ mesons with the nucleon, viz.
\begin{eqnarray}\label{DNInt}
B^-(b\overline{u})+p(uud)~or~n(udd)&\rightarrow&
\Lambda_b^0(udb)~or~\Sigma_b^-(ddb),
\nonumber \\
D^0(c\overline{u})+p(uud)~or~n(udd)&\rightarrow& \Lambda_c^+,
\Sigma_c^+(udc)~or~\Sigma_c^0(ddc).
\end{eqnarray}
Hence, we need to know the exact values of the strong coupling constants $g_{\Lambda_bNB}$, $g_{\Lambda_cND}$,  $g_{\Sigma_bNB}$ and $g_{\Sigma_cND}$ entering the Born term in the calculations
\cite{Azizi1,Kumar,Wang1,Wang2,Hayashigaki}. Note that, among these couplings, we have only one approximate prediction for the strong coupling $g_{\Lambda_cND}$ 
in the literature calculated at zero transferred momentum square taking
the Borel masses in the initial and final channels as the same \cite{Navarra}. We shall also refer to a pioneering work \cite{Choe}, which  estimates the strong coupling constant 
$g_{NK\Lambda}$. Here we should also stress that our work on the calculation of the strong coupling constants   $g_{\Sigma_bNB}$ and $g_{\Sigma_cND}$ is in progress. 

The layout of this article is as follows. The next section presents the details of the calculations
of the strong coupling constants under consideration. In section 3, we  numerically analyze the sum rules obtained and discuss the results.

\section{ The strong coupling form factors}
The purpose of the present section is to give the details of the
calculations of the coupling form factors $g_{\Lambda_b NB}(q^2)$ and $g_{\Lambda_cND}(q^2)$. The values of these form factors at $Q^2=-q^2=-m_{B[D]}^2$ give the strong coupling constants among the participating particles. To fulfill this aim, the starting point is the usage of the
following three-point correlation function:
\begin{eqnarray}\label{CorrelationFunc1}
\Pi(p,p^{\prime},q)=i^2 \int d^4x~ \int d^4y~e^{-ip\cdot x}~
e^{ip^{\prime}\cdot y}~{\langle}0| {\cal T}\left (
J_{N}(y)~J_{B[D]}(0)~
\bar{J}_{\Lambda_b[\Lambda_c]}(x)\right)|0{\rangle},
\end{eqnarray}
where ${\cal T}$ denotes the time ordering operator and $q=p-p'$ is
transferred momentum. The three-point correlation function contains
interpolating currents that can be written in terms of the quark
field operators as:
\begin{eqnarray}\label{InterpolatingCurrents}
J_{\Lambda_{b}[\Lambda_{c}]}(x)&=&\varepsilon_{abc}u^{a^T}(x)C\gamma_5d^{b}(x)b[c]^{c}(x),
\nonumber \\
J_{N}(y)&=&\varepsilon_{ijk}\Big(u^{i^T}(y)C\gamma_{\mu}u^{j}(y)\Big)\gamma_{5}\gamma_{\mu}d^{k}(y),
\nonumber \\
J_{B[D]}(0)&=&\bar{u}(0)\gamma_5b[c](0),
\end{eqnarray}
where $C$ is the charge conjugation operator.

The calculation of the three-point correlation function is made via
following two different ways. In the first way, which is called as
hadronic side, one calculates it in terms of the hadronic degrees of
freedom. In the second way, which is called as OPE side,  it is calculated
in terms of quark and gluon degrees of freedom using the
operator product expansion in deep Euclidean region. These two sides are then matched  to obtain the QCD sum rules
for the coupling form factors. We apply a double Borel
transformation with respect to the variables $p^2$ and $p'^2$ to both sides to  suppress the contributions of the higher states and continuum.

The calculation of the hadronic side of the correlation function requires its saturation  with complete sets of appropriate
$\Lambda_{b}[\Lambda_{c}]$, $B[D]$ and $N$ hadronic states having
the same quantum numbers as their interpolating currents. This step is followed by performing the four-integrals
over $x$ and $y$, which leads to
\begin{eqnarray} \label{physide1}
\Pi(p,p^{\prime},q)&=&\frac{\langle 0 \mid
 J_{N}\mid N(p^{\prime})\rangle \langle 0 \mid
 J_{B[D]}\mid B[D](q)\rangle \langle
\Lambda_{b}[\Lambda_{c}](p) \mid
 \bar{J}_{\Lambda_{b}[\Lambda_{c}]}\mid 0\rangle }{(p^2-m_{\Lambda_{b}[\Lambda_{c}]}^2)(p^{\prime^2}-m_{N}^2)(q^2-m_{B[D]}^2)}
 \nonumber \\
&\times&\langle N(p^{\prime})B[D](q)\mid
\Lambda_{b}[\Lambda_{c}](p)\rangle+\cdots~,
\end{eqnarray}
where $\cdots$ represents the contributions coming from the higher states and continuum. The matrix elements in this equation
are parameterized as 
\begin{eqnarray}\label{matriselement1}
\langle 0 \mid
 J_{N}\mid N(p^{\prime})\rangle&=&\lambda_N u_N(p^{\prime}, s^{\prime}),
\nonumber \\
\langle \Lambda_b(p) \mid
 \bar{J}_{\Lambda_b[\Lambda_c]}\mid
 0\rangle&=&\lambda_{\Lambda_b[\Lambda_c]}\bar{u}_{\Lambda_b[\Lambda_c]}(p, s),
\nonumber \\
\langle 0 \mid
 J_{B[D]}\mid B[D](q)\rangle&=&i\frac{m_{B[D]}^2f_{B[D]}}{m_u+m_{b[c]}},
\nonumber \\
\langle N(p^{\prime})B[D](q)\mid
 \Lambda_b[\Lambda_c](p)\rangle&=&g_{\Lambda_bNB[\Lambda_c ND]}\bar{u}_N(p^{\prime},s^{\prime})i
 \gamma_5 u_{\Lambda_b[\Lambda_c]}(p, s),
\end{eqnarray}
where  $\lambda_N$ and $\lambda_{\Lambda_b[\Lambda_c]}$ are the residues; and $u_N$ and 
$u_{\Lambda_b[\Lambda_c]}$ are the spinors for the nucleon and
$\Lambda_b[\Lambda_c]$ baryon, respectively. In the above equations, $f_{B[D]}$ is the
leptonic decay constant of $B[D]$ meson and
$g_{\Lambda_bNB[\Lambda_cND]}$ is the strong  coupling form factor among $\Lambda_b[\Lambda_c]$, $N$ and $B[D]$ particles. The use of
Eqs.~(\ref{matriselement1}) in Eq.~(\ref{physide1}) is followed by
summing over the spins of the $N$ and $\Lambda_b[\Lambda_c]$
baryons, i.e.
\begin{eqnarray}\label{spinor}
\sum_{s'}u_{N}(p^{\prime},s^{\prime})\bar{u}_{N}(p^{\prime},s^{\prime})&=&
\not\!p^{\prime}+m_{N}~,
\nonumber \\
\sum_{s}u_{\Lambda_b[\Lambda_c]}(p,
s)\bar{u}_{\Lambda_b[\Lambda_c]}(p, s)&=&
\not\!p+m_{\Lambda_b[\Lambda_c]}~.
\end{eqnarray}
As a result, we have
\begin{eqnarray} \label{physide1Last}
\Pi(p,p^{\prime},q)&=&i^2\frac{m_{B[D]}^2f_{B[D]}}{m_{b[c]}+m_u}\frac{\lambda_N\lambda_{\Lambda_b[\Lambda_c]}
g_{\Lambda_bNB[\Lambda_c
ND]}}{(p^2-m_{\Lambda_b[\Lambda_c]}^2)(p^{\prime^2}-m_N^2)(q^2-m_{B[D]}^2)}
\nonumber \\
&\times&\Big\{(m_Nm_{\Lambda_b[\Lambda_c]}-
m_{\Lambda_b[\Lambda_c]}^2)\gamma_5+(m_{\Lambda_b[\Lambda_c]}-m_N)\not\!p\gamma_5+\not\!q\not\!p\gamma_5-m_{\Lambda_b[\Lambda_c]}
\not\!q\gamma_5\Big\}\nonumber\\
&+&\cdots~.
\end{eqnarray}
The final form of the hadronic side of the correlation function is
obtained after the application of the double Borel transformation
with respect to the initial and final momenta squared, viz.
\begin{eqnarray} \label{physide1Last1}
\widehat{\textbf{B}}\Pi(q)&=&i^2\frac{m_{B[D]}^2f_{B[D]}}{m_{b[c]}+m_u}\frac{\lambda_N\lambda_{\Lambda_b[\Lambda_c]}
g_{\Lambda_bNB[\Lambda_cND]}}{(q^2-m_{B[D]}^2)}e^{-\frac{m_{\Lambda_b[\Lambda_c]}^2}{M^2}}
e^{-\frac{m_N^2}{M^{\prime^2}}}
\nonumber \\
&\times&\Big\{(m_Nm_{\Lambda_b[\Lambda_c]}-
m_{\Lambda_b[\Lambda_c]}^2)\gamma_5+(m_{\Lambda_b[\Lambda_c]}-m_N)\not\!p\gamma_5+\not\!q\not\!p\gamma_5-m_{\Lambda_b[\Lambda_c]}
\not\!q\gamma_5\Big\}\nonumber\\&+&\cdots~,
\end{eqnarray}
where $M^2$ and $M^{\prime^2}$ are Borel mass parameters.

The OPE side of the  correlation function is calculated
in deep Euclidean region, where $p^2\rightarrow -\infty$ and
$p'^2\rightarrow -\infty$. To proceed, the explicit expressions of
the interpolating currents are inserted into the correlation
function in Eq.~(\ref{CorrelationFunc1}). After contracting out all
quark pairs via Wick's theorem we get
\begin{eqnarray}\label{correlfuncOPE1}
\Pi(p, p^{\prime}, q)&=&i^2\int d^{4}x\int d^{4}ye^{-ip\cdot
x}e^{ip^{\prime}\cdot y}\varepsilon_{abc}\varepsilon_{ij\ell}
\nonumber \\
&\times&
\Bigg\{\gamma_5\gamma_{\mu}S^{cj}_{d}(y-x)\gamma_{5}CS_{u}^{bi^T}(y-x)C\gamma_{\mu}S^{ah}_{u}(y)
\gamma_{5}S_{b[c]}^{h\ell}(-x)
\nonumber \\
&-&\gamma_5\gamma_{\mu}S^{cj}_{d}(y-x)\gamma_{5}CS_{u}^{ai^T}(y-x)C\gamma_{\mu}S^{bh}_{u}(y)
\gamma_{5}S_{b[c]}^{h\ell}(-x)
 \Bigg\}~,
\end{eqnarray}
where $ S^{i\ell}_{b[c]}(x)$ represents the heavy quark propagator
which is given by \cite{Reinders}
\begin{eqnarray}\label{heavypropagator}
S_{b[c]}^{i\ell}(x)&=&\frac{i}{(2\pi)^4}\int d^4k e^{-ik \cdot x}
\left\{ \frac{\delta_{i\ell}}{\!\not\!{k}-m_{b[c]}}
-\frac{g_sG^{\alpha\beta}_{i\ell}}{4}\frac{\sigma_{\alpha\beta}(\!\not\!{k}+m_{b[c]})+
(\!\not\!{k}+m_{b[c]})\sigma_{\alpha\beta}}{(k^2-m_{b[c]}^2)^2}\right.\nonumber\\
&&\left.+\frac{\pi^2}{3} \langle \frac{\alpha_sGG}{\pi}\rangle
\delta_{i\ell}m_{b[c]}
\frac{k^2+m_{b[c]}\!\not\!{k}}{(k^2-m_{b[c]}^2)^4} +\cdots\right\}
\, ,
\end{eqnarray}
and $S_{u}(x)$ and $S_d(x)$ are the light quark propagators and
are given by
\begin{eqnarray}\label{lightpropagator}
S_{q}^{ij}(x)&=& i\frac{\!\not\!{x}}{ 2\pi^2 x^4}\delta_{ij}
-\frac{m_q}{4\pi^2x^2}\delta_{ij}-\frac{\langle
\bar{q}q\rangle}{12}\Big(1 -i\frac{m_q}{4}
\!\not\!{x}\Big)\delta_{ij} -\frac{x^2}{192}m_0^2\langle
\bar{q}q\rangle\Big(1-i\frac{m_q}{6} \!\not\!{x}\Big)\delta_{ij}
\nonumber \\
&-&\frac{ig_s
G_{\theta\eta}^{ij}}{32\pi^2x^2}\big[\!\not\!{x}\sigma^{\theta\eta}+\sigma^{\theta\eta}\!\not\!{x}\big]
+\cdots \, .
\end{eqnarray}
The substitution of these explicit forms of the heavy and light
quark propagators into Eq.~(\ref{correlfuncOPE1}) is followed by
the usage of the following Fourier transformations in $D=4$ dimensions:
\begin{eqnarray}\label{intyx}
\frac{1}{[(y-x)^2]^n}&=&\int\frac{d^Dt}{(2\pi)^D}e^{-it\cdot(y-x)}~i~(-1)^{n+1}~2^{D-2n}~\pi^{D/2}~
\frac{\Gamma(D/2-n)}{\Gamma(n)}\Big(-\frac{1}{t^2}\Big)^{D/2-n},
\nonumber \\
\frac{1}{[y^2]^n}&=&\int\frac{d^Dt^{\prime}}{(2\pi)^D}e^{-it^{\prime}\cdot
y}~i~(-1)^{n+1}~2^{D-2n}~\pi^{D/2}~
\frac{\Gamma(D/2-n)}{\Gamma(n)}\Big(-\frac{1}{t^{\prime^2}}\Big)^{D/2-n}.
\end{eqnarray}
Then, the four-$x$ and four-$y$ integrals are performed in the sequel of
the replacements $x_{\mu}\rightarrow i\frac{\partial}{\partial
p_{\mu}}$ and
 $y_{\mu}\rightarrow -i\frac{\partial}{\partial p'_{\mu}}$. As a result,  these integrals turn into Dirac delta functions which are used to take the four-integrals over $k$ and
 $t^{\prime}$. Finally  the Feynman parametrization and

\begin{eqnarray}\label{Int}
\int d^4t\frac{(t^2)^{\beta}}{(t^2+L)^{\alpha}}=\frac{i \pi^2
(-1)^{\beta-\alpha}\Gamma(\beta+2)\Gamma(\alpha-\beta-2)}{\Gamma(2)
\Gamma(\alpha)[-L]^{\alpha-\beta-2}},
\end{eqnarray}
are used to perform the remaining four-integral over $t$.

 The correlation function in OPE side is obtained in terms of
different structures as
\begin{eqnarray}\label{correlfuncOPE1Last}
\Pi(p, p^{\prime},
q)&=&\Pi_1(q^2)\gamma_5+\Pi_2(q^2)\!\not\!{p}\gamma_5+\Pi_3(q^2)\!\not\!{q}\!\not\!{p}\gamma_5
+\Pi_4(q^2)\!\not\!{q}\gamma_5,
\end{eqnarray}
where each $\Pi_i(q^2)$  function includes the
contributions coming from both the perturbative and non-perturbative
parts and can be written as
\begin{eqnarray}\label{QCDside1}
\Pi_i(q^2)=\int^{}_{}ds\int^{}_{}ds^{\prime}
\frac{\rho_i^{pert}(s,s^{\prime},q^2)+\rho_i^{non-pert}(s,s^{\prime},q^2)}{(s-p^2)
(s^{\prime}-p^{\prime^2})}~.
\end{eqnarray}
The imaginary parts of the $\Pi_{i}$ functions give the spectral
densities $\rho_i(s,s',q^2)$ appearing in the last equation, viz.
$\rho_i(s,s',q^2)=\frac{1}{\pi}Im[\Pi_{i}]$. As examples, we present only the explicit forms of the spectral functions 
$\rho_1^{pert}(s,s^{\prime},q^2)$ and
$\rho_1^{non-pert}(s,s^{\prime},q^2)$ corresponding to the
Dirac structure $\gamma_5$, which are obtained as
\begin{eqnarray}\label{rho1pert}
\rho_1^{pert}(s,s^{\prime},q^2)&=&
\Bigg\{-\frac{m_{b[c]}m_us^{\prime^2}}{64\pi^4(q^2-m_{b[c]}^2)}
\Theta[L_1(s,s^{\prime},q^2)]+\int_{0}^{1}dx \int_{0}^{1-x}dy
\frac{1}{64\pi^4u^3}
\nonumber \\
&\times&\Big[2m_{b[c]}^4x^2\Big(1+3x^2-y+6xy-4x)\Big)+m_{b[c]}^3x\Big(3m_d
u(2x-1) +m_u(3+2x^2
\nonumber \\
&-&3y-5x-2xy)\Big)+2m_{b[c]}^2x\Big(s(12x^4+y^2-y-30x^3+36x^3y-6x+20xy
\nonumber \\
&-&13xy^2+24x^2
-55x^2y+24x^2y^2)+q^2xy(18x-24xy+7y-12x^2-6)
\nonumber \\
&+&s^{\prime}y(12x^3+7y-4y^2-27x^2+36x^2y+18x-43xy+24xy^2-3) \Big)
\nonumber \\
&+&2s^2u^2x\Big(10x^3+6x-15xy+2y-16x^2+20x^2y\Big)+2q^4x^2y^2\Big(10x^2-7y
\nonumber \\
&-&16x+20xy+6\Big)+2s^{\prime^2}y^2u^2
\Big(10x^2-3y-12x+20xy\Big)-4q^2s^{\prime}xy^2\Big(10x^3
\nonumber \\
&+&9y-5y^2-24x^2+30x^2y+18x-39xy+20xy^2-4\Big)+2suy
\Big(q^2x(32x^2
\nonumber \\
&-&40x^2y-20x^3-2y-13x+22xy+1)+s^{\prime}(20x^4-48x^3+60x^3y-y+y^2
\nonumber \\
&-&8x+27xy-18xy^2 +36x^2-86x^2y+40x^2y^2)\Big)+3m_{b[c]}m_u
u\Big(q^2x(x+2y
\nonumber \\
&-&3xy-1)+sux(3x-1)+s^{\prime}u(3xy-x-y)\Big)
-m_{b[c]}m_u\Big(q^2x(3x^2y-3x^2
\nonumber \\
&+&7y-4y^2+6x-10xy-3xy^2-3)-sux(3x^2-y-6x-6xy+3)
\nonumber \\
&-&3s^{\prime}u(x^2y-x^2+y -y+x-3xy-xy^2)\Big)
\Big]\Theta[L_2(s,s^{\prime},q^2)] \Bigg\},
\end{eqnarray}
and
\begin{eqnarray}\label{rho1nonpert}
\rho_1^{non-pert}(s,s^{\prime},q^2)&=&\Big\{\frac{1}{16\pi^2(m_{b[c]}^2-q^2)}\Big[2m_{b[c]}m_dm_u\langle
\bar{d}d\rangle+\Big(m_{b[c]}(3m_u^2-3m_dm_u-2s^{\prime})
\nonumber \\
&+&m_d(4m_u^2+s-s^{\prime})+2m_us^{\prime} \Big)\langle
\bar{u}u\rangle\Big]-\langle
\alpha_s\frac{G^2}{\pi}\rangle\Bigg[\frac{m_{b[c]}m_uq^2s^{\prime^2}}{192\pi^2(q^2-m_{b[c]}^2)^4}
\nonumber \\
&-&\frac{9m_{b[c]}s^{\prime}(m_d+m_u)+2s^{\prime}(q^2-2s+5s^{\prime})}{1152\pi^2(q^2-m_{b[c]}^2)^2}-
\frac{m_{b[c]}(m_d-3m_u)}{128\pi^2(q^2-m_{b[c]}^2)}\Bigg]
\nonumber \\
&-&m_0^2\langle \bar{d}d\rangle
\frac{3m_{b[c]}+4m_d}{96\pi^2(m_{b[c]}^2-q^2)}+m_0^2\langle
\bar{u}u\rangle
\frac{9m_{b[c]}+3m_d-7m_u}{96\pi^2(m_{b[c]}^2-q^2)}\Bigg\}\Theta[L_1(s,s^{\prime},q^2)]
\nonumber \\
 &+&\int_{0}^{1}dx
\int_{0}^{1-x}dy\Bigg\{\frac{1}{8\pi^2u}\Big[\langle
\bar{d}d\rangle\Big(m_{b[c]}-2m_{b[c]}x-m_uu+m_d(3x-1)(y+u)\Big)
\nonumber \\
&+&\langle
\bar{u}u\rangle\Big(m_{b[c]}-2m_{b[c]}x-4m_du-2m_u(y-3xy-3xu)\Big)\Big]
+\langle \alpha_s\frac{G^2}{\pi}\rangle  \frac{1}{96\pi^2u^3}
\nonumber
\\&\times&\Big[3u^2(3x-1)(y+u)+xy(1-y+x(3x+6y-4))\Big]
\Bigg\}\Theta[L_2(s,s^{\prime},q^2)], \nonumber \\
\end{eqnarray}
where
\begin{eqnarray}\label{teta1}
L_1(s,s^{\prime},q^2)&=&s^{\prime},
\nonumber \\
L_2(s,s^{\prime},q^2)&=&-m_{b[c]}^2x+sx-sx^2+s^{\prime}y+q^2xy-sxy-s^{\prime}xy-s^{\prime}y^2,
\nonumber \\
u&=&x+y-1,
\end{eqnarray}
 with $\Theta[...]$ being the unit-step function.

As we previously mentioned, the QCD sum rules for the strong form factors are  obtained by matching the hadronic and OPE sides of the correlation function. As a result,  for $\gamma_5$ structure, we get
\begin{eqnarray}\label{couplingconstant}
&&g_{\Lambda_{b}NB[\Lambda_{c}ND]}(q^2)=-e^{\frac{m_{\Lambda_b[\Lambda_c]}^2}{M^2}}e^{\frac{m_N^2}{M^{\prime^2}}}~
\frac{(m_{b[c]}+m_u)(q^2-m_{B[D]}^2)}{m_{B[D]}^2f_{B[D]}\lambda_{\Lambda_{b}[\Lambda_c]}^{\dag}\lambda_N(m_Nm_{\Lambda_{b}[\Lambda_c]}-
m_{\Lambda_b[\Lambda_c]}^2)}
\nonumber \\
&\times&
\Bigg\{\int^{s_0}_{(m_{b[c]}+m_{u}+m_{d})^2}ds\int^{s'_0}_{(2m_u+m_d)^2}ds^{\prime}e^{-\frac{s}{M^2}}
e^{-\frac{s^{\prime}}{M^{\prime^2}}}
\Big[\rho_1^{pert}(s,s^{\prime},q^2)+\rho_1^{non-pert}(s,s^{\prime},q^2)\Big]\Bigg\}~,\nonumber\\
\end{eqnarray}
where $s_0$ and $s'_0$ are continuum thresholds in
$\Lambda_b[\Lambda_c]$ and $N$ channels, respectively.

\section{Numerical results}

This section contains the numerical analysis of the obtained sum
rules for the strong coupling form factors including their behavior in terms of  $Q^2=-q^2$. For the analysis, we use the input
parameters given in  table 1.
\begin{table}[ht]\label{Table1}
\centering \rowcolors{1}{lightgray}{white}
\begin{tabular}{cc}
\hline \hline
   Parameters  &  Values
           \\
\hline \hline
$m_{b}$              & $(4.18\pm0.03)~\mbox{GeV}$\cite{Olive}\\
$m_{c}$              & $(1.275\pm0.025)~\mbox{GeV}$\cite{Olive}\\
$m_{d}$              & $4.8^{+0.5}_{-0.3}~\mbox{MeV}$\cite{Olive}\\
$ m_{u} $            &$2.3^{+0.7}_{-0.5}~\mbox{MeV}$ \cite{Olive}\\
$ m_{B}$    &   $ (5279.26\pm0.17)~\mbox{MeV}$ \cite{Olive}  \\
$ m_{D}$    &   $ (1864.84\pm0.07)~\mbox{MeV}$ \cite{Olive}  \\
$ m_{N} $      &   $ (938.272046\pm0.000021)~\mbox{MeV}$  \cite{Olive} \\
$ m_{\Lambda_b} $      &   $ (5619.5\pm0.4) ~\mbox{MeV} $ \cite{Olive}  \\
$ m_{\Lambda_c} $      &   $(2286.46\pm 0.14) ~\mbox{MeV} $ \cite{Olive}  \\
$ f_{B} $      &   $(248\pm23_{exp}\pm25_{Vub}) ~\mbox{MeV}$
\cite{Khodjamirian1} \\
$ f_{D} $      &   $(205.8\pm8.5\pm2.5) ~\mbox{MeV}$  \cite{CLEO} \\
$ \lambda_{N}^2 $      &   $0.0011\pm0.0005  ~\mbox{GeV$^6$}$  \cite{Azizi2} \\
$ \lambda_{\Lambda_b} $      &   $(3.85\pm0.56)10^{-2}$ $\mbox{GeV$^3$}$  \cite{Azizi} \\
$ \lambda_{\Lambda_c} $      &   $(3.34\pm0.47)10^{-2}$ $\mbox{GeV$^3$}$  \cite{Azizi} \\
$  \langle \bar{u}u\rangle(1~GeV)=\langle \bar{d}d\rangle(1~GeV)$&
$-(0.24\pm0.01)^3 $ $\mbox{GeV$^3$}$
 \cite{Ioffe} \\
$ \langle\frac{\alpha_sG^2}{\pi}\rangle $       &   $(0.012\pm0.004)$ $~\mbox{GeV$^4$}$
\cite{belyaev}   \\
$ m_0^2(1~GeV) $       & $(0.8\pm0.2)$ $~\mbox{GeV$^2$}$
\cite{belyaev}   \\
 \hline \hline
\end{tabular}
\caption{Input parameters used in  calculations.}
\end{table}

The analysis starts by the determination of the working regions for
the auxiliary parameters $M^2$, $M'^2$, $s_0$ and $s'_0$. These
parameters, which arise due to the double Borel transformation and continuum subtraction, are not physical parameters so  the strong coupling form factors should be
almost independent of these parameters. Being related to the energy
of the first excited states in the initial and final channels, the continuum thresholds are not
completely arbitrary. The continuum thresholds $s_{0}$ and $s'_0$ are the energy squares which characterize the beginning of the continuum. If we denote the ground states masses 
 in the initial and final channels respectively by $m$ and $m'$,  the quantities $\sqrt{s_0}-m$ and $\sqrt{s'_0}-m'$
are the energies needed  to excite the particles to their first excited states  with the same quantum numbers. The $\sqrt{s_0}-m$ and $\sqrt{s'_0}-m'$ are well known for the states under consideration 
\cite{Olive}, where they lie roughly  between $0.1~\mbox{GeV}$ and $0.3~\mbox{GeV}$. These values lead  to the working
intervals of the continuum thresholds as $32.7[5.7]~\mbox{GeV$^2$}\leq
s_0\leq34.5[6.7]~\mbox{GeV$^2$}$ and $1.08~\mbox{GeV$^2$}\leq
s'_0\leq1.56~\mbox{GeV$^2$}$ for the strong vertex
$\Lambda_bNB[\Lambda_cND]$.

In the determination of the working regions
of Borel parameters $M^2$ and $M'^2$, one considers the pole dominance
as well as the convergence of the OPE. In technique language,  the upper bounds on these parameters are obtained by
requiring  that the pole contribution 
exceeds   the contributions of the higher states and continuum,
i.e. the condition
\bea
\label{nolabel}
{\ds \int^{\infty}_{s_0}ds\int^{\infty}_{s'_0}ds^{\prime}e^{-\frac{s}{M^2}}
e^{-\frac{s^{\prime}}{M^{\prime^2}}}\rho_i(s,s^{\prime},Q^2) \over \ds 
\int^{\infty}_{s_{min}}ds\int^{\infty}_{s'_{min}}ds^{\prime}e^{-\frac{s}{M^2}}
e^{-\frac{s^{\prime}}{M^{\prime^2}}}\rho_i(s,s^{\prime},Q^2)}
< 1/3~,
\eea
should be satisfied, where for each structure $\rho_i(s,s^{\prime},Q^2)=\rho_i^{pert}(s,s^{\prime},Q^2)+\rho_i^{non-pert}(s,s^{\prime},Q^2)$, $s_{min}=(m_{b[c]}+m_{u}+m_{d})^2$ and $s'_{min}=(2m_u+m_d)^2$. The lower bounds on $M^2$ and $M'^2$  are
obtained by demanding that the contribution of the perturbative
part exceeds   the non-perturbative contributions.
These considerations 
  lead to the windows
$10[2]~\mbox{GeV$^2$}\leq M^2\leq 20[6]~\mbox{GeV$^2$}$ and
$1~\mbox{GeV$^2$}\leq M'^2\leq 3~\mbox{GeV$^2$}$ for the Borel mass
parameters corresponding to the strong vertex
$\Lambda_bNB[\Lambda_cND]$ in which our results have  weak dependencies on the Borel
mass parameters (see figures 1-2).

\begin{figure}[h!]
\includegraphics[totalheight=6cm,width=8cm]{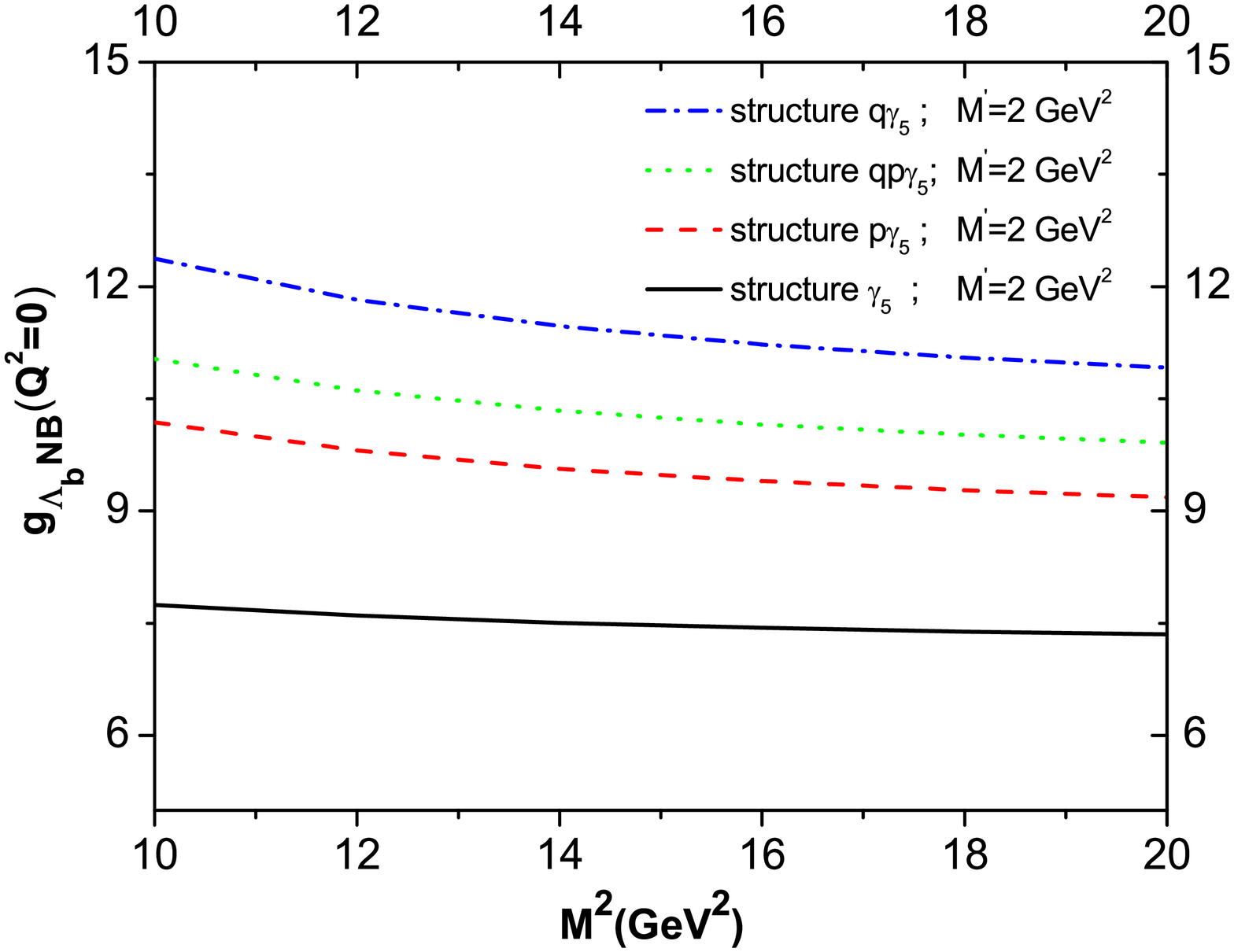}
\includegraphics[totalheight=6cm,width=8cm]{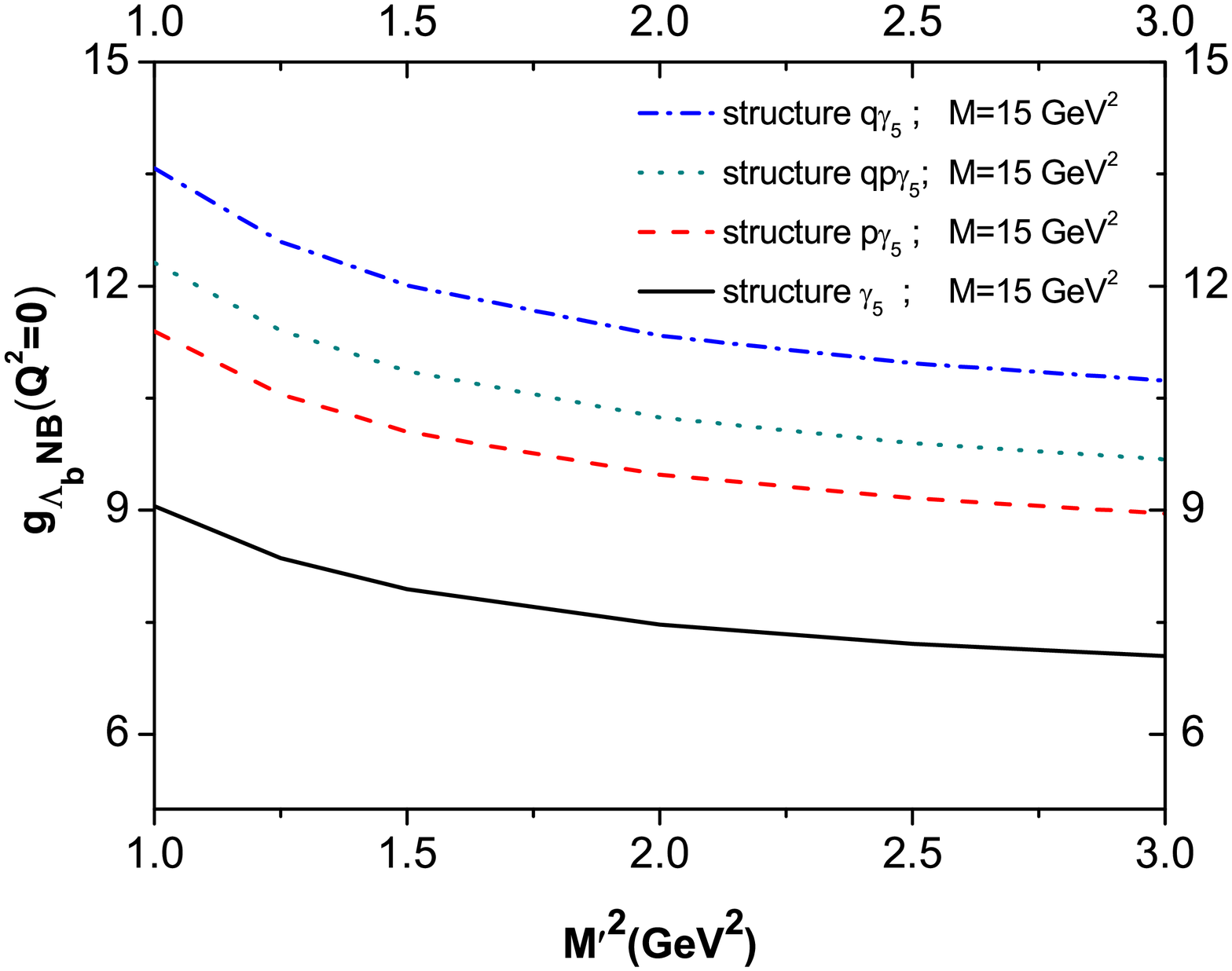}
\caption{\textbf{Left:} $g_{\Lambda_bNB}(Q^2=0)$ as a function of
the Borel mass $M^2$ at average values of continuum thresholds. \textbf{Right:}
 $g_{\Lambda_bNB}(Q^2=0)$ as a function of the
Borel mass $M^{\prime^2}$ at average values of continuum thresholds. } \label{gLamdabNBMsqMpsq}
\end{figure}
\begin{figure}[h!]
\includegraphics[totalheight=6cm,width=8cm]{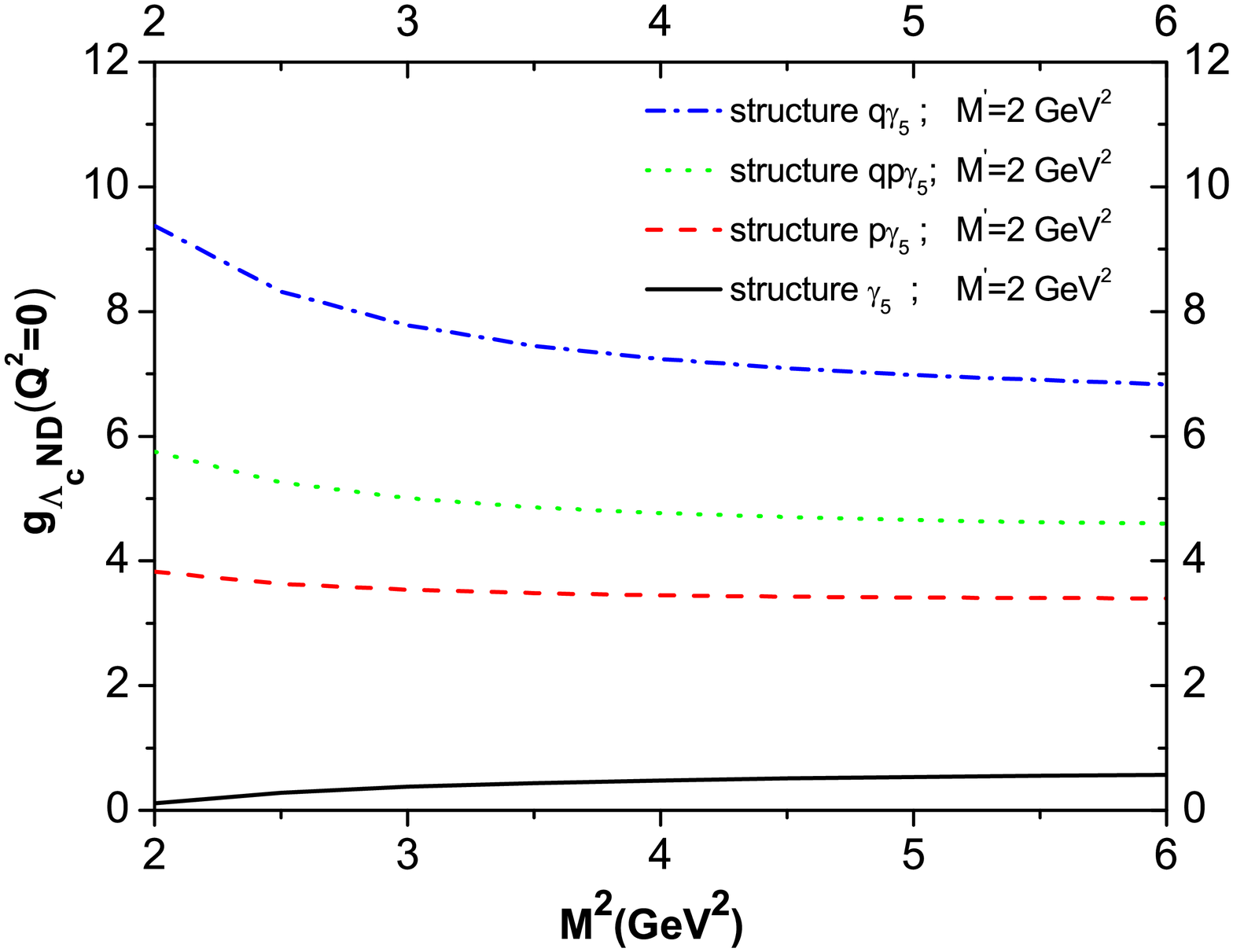}
\includegraphics[totalheight=6cm,width=8cm]{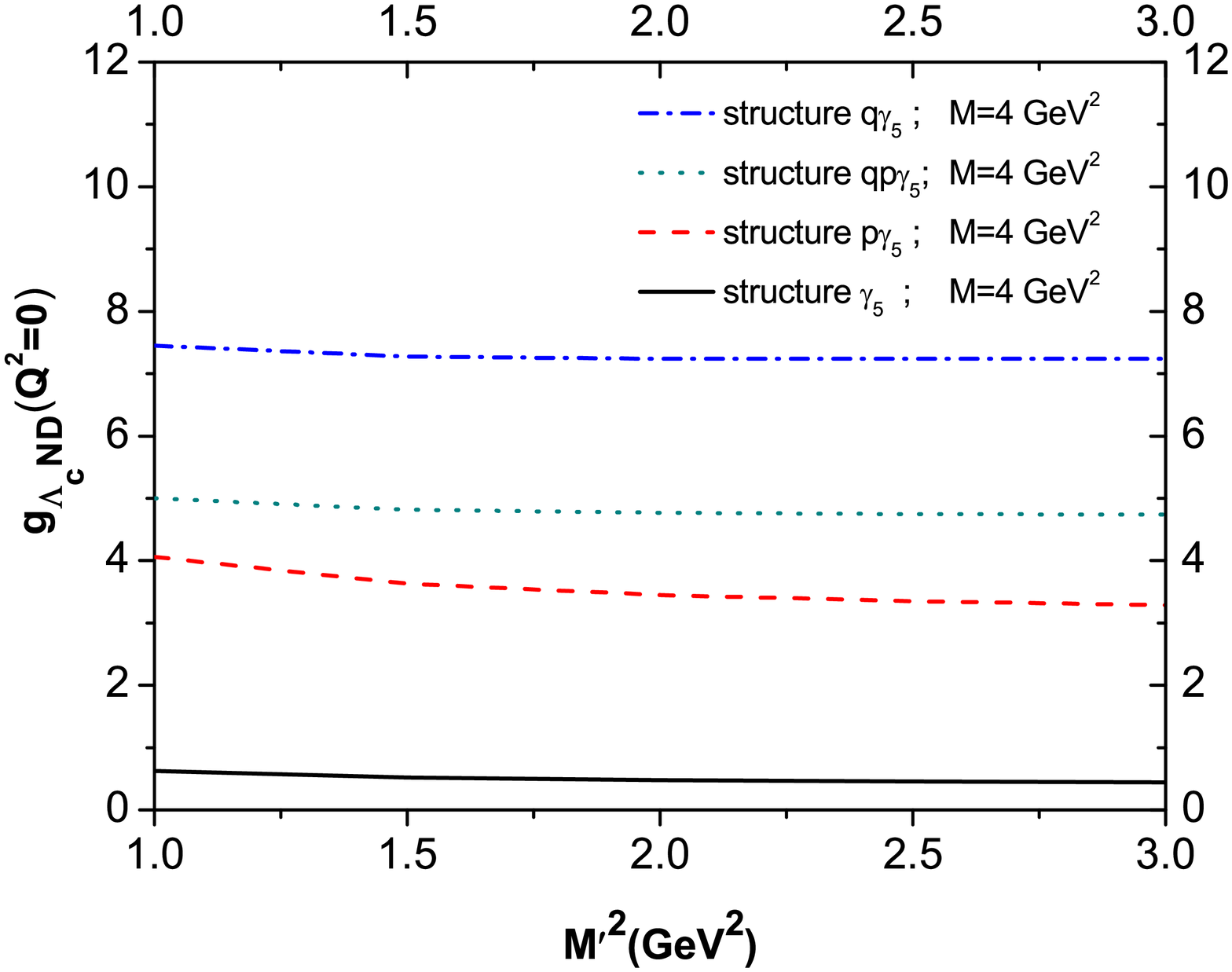}
\caption{The same as figure 1 but for $g_{\Lambda_cND}(Q^2=0)$. } \label{gLamdabNBMsqMpsq}
\end{figure}

\begin{figure}[h!]
\includegraphics[totalheight=6cm,width=8cm]{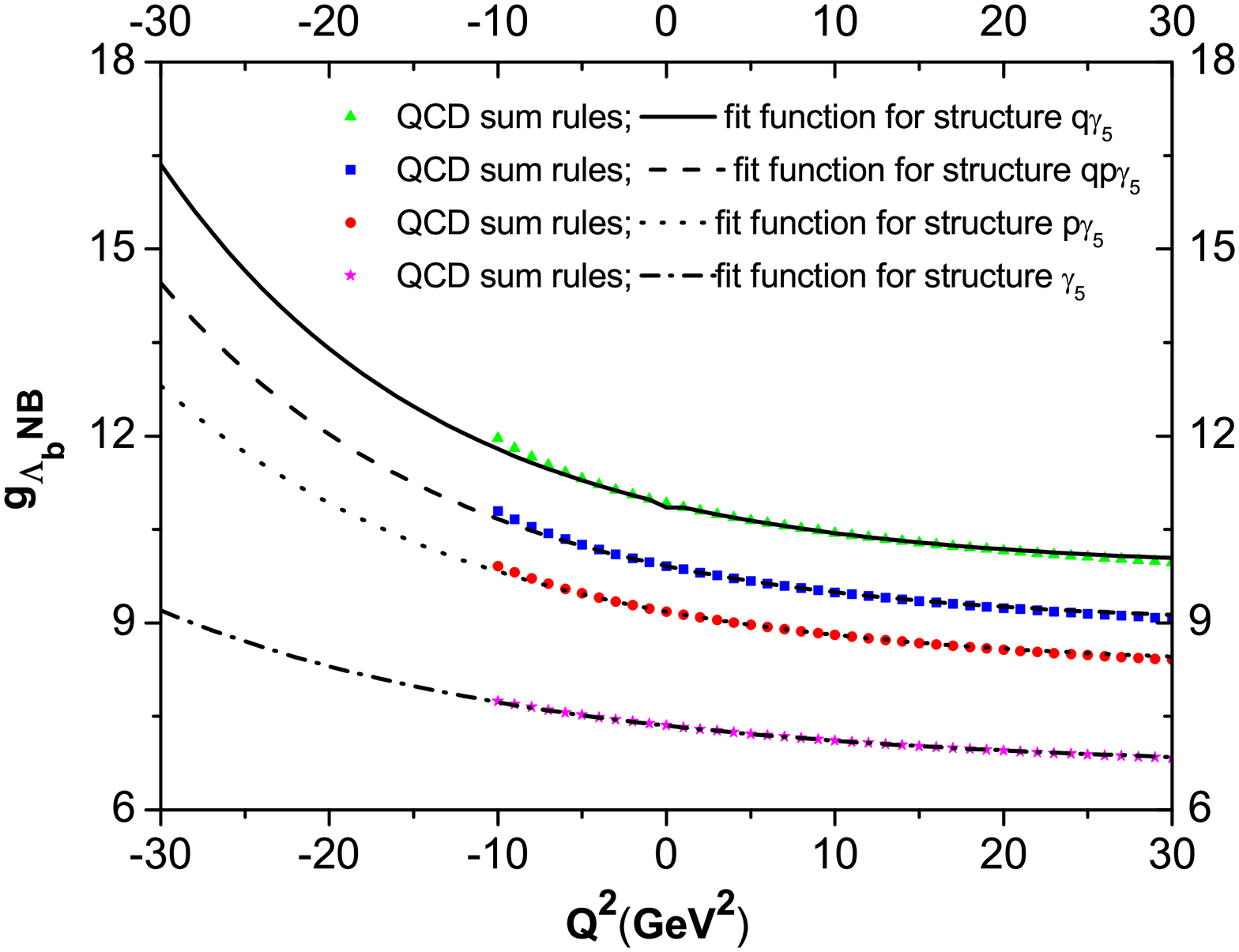}
\includegraphics[totalheight=6cm,width=8cm]{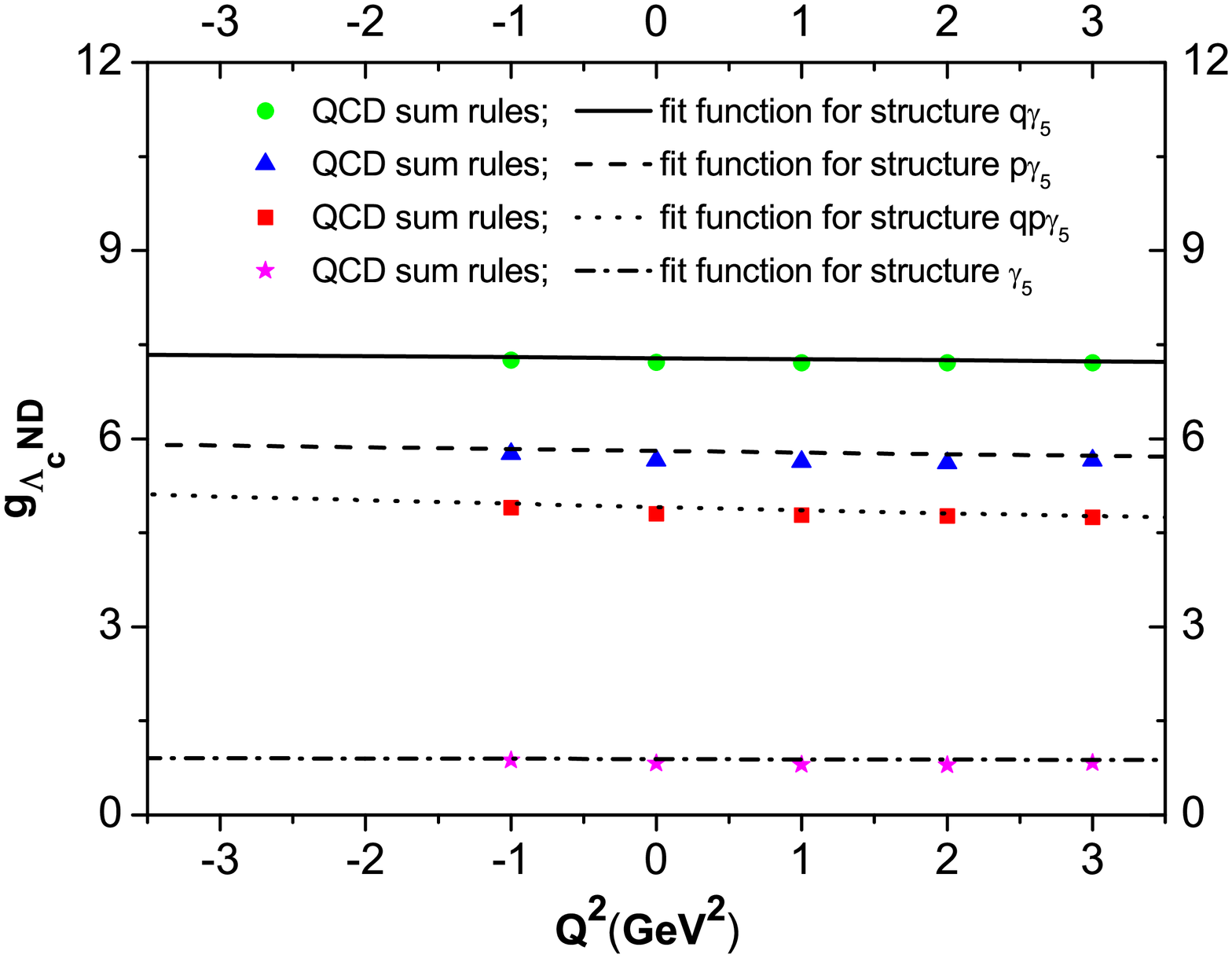}
\caption{\textbf{Left:} $g_{\Lambda_bNB}(Q^2)$ as a function of
 $Q^2$ at average values of the continuum thresholds and Borel mass parameters. \textbf{Right:}
 $g_{\Lambda_cND}(Q^2)$ as a function of  $Q^2$ at average values of the continuum thresholds and Borel mass parameters. } \label{gLamdabNBMsqMpsq}
\end{figure}

Now, we use the working regions of auxiliary parameters as well as values of other input parameters to find out the dependency of the strong coupling form factors on
$Q^2$.  Our numerical calculations reveal that  the following fit function well  describes the strong
coupling form factors in terms of $Q^2$:
\begin{eqnarray}\label{fitfunc}
g_{\Lambda_bNB[\Lambda_cND]}(Q^2)=c_1\exp\Big[-\frac{Q^2}{c_2}\Big]+c_3,
\end{eqnarray}
where the values of the parameters $c_1$, $c_2$ and $c_3$ for
different structures are presented in  tables \ref{fitparam} and
\ref{fitparam1} for $\Lambda_bNB$ and $\Lambda_cND$, respectively.  In figure 3, we depict the dependence of the strong coupling form factors on $Q^2$ at average values of the continuum thresholds and
Borel mass parameters for both the QCD sum rules and fitting results. From this figure, we see that the QCD sum rules are truncated at some points at negative values of $Q^2$ and the fitting results coincide well
with the sum rules predictions up to these points.
The values of the strong coupling constants obtained from the fit function at $Q^2=-m_{B[D]}^2$ for all structures are given in table \ref{couplingconstant}. The errors appearing in
the results are due to the uncertainties of the input parameters and
those  coming from the calculations of the working regions for the auxiliary parameters. From table 4, we see that all structures except that $\gamma_5$ lead to very close results. We also depict the average
of the coupling constants under consideration, obtained from all the structures used,  in table 4.
\begin{table}[h]
\renewcommand{\arraystretch}{1.5}
\addtolength{\arraycolsep}{3pt}
$$
\begin{array}{|c|c|c|c|}
\hline \hline
       \mbox{structure}    & c_1 & c_2 (\mbox{GeV$^2$})&c_3     \\
\hline
  \mbox{$\gamma_5$} &0.69\pm 0.21&22.96\pm6.66&6.67\pm2.00 \\
  \hline
  \mbox{$\!\not\!{p}\gamma_5$} &0.90\pm 0.27&18.60\pm5.58&8.28\pm2.32 \\
  \hline
   \mbox{$\!\not\!{q}\gamma_5$} &1.04\pm 0.31&16.40\pm4.92&9.87\pm2.67 \\
  \hline
  \mbox{$\!\not\!{q}\!\not\!{p}\gamma_5$} &0.95\pm 0.28&17.10\pm4.79&8.96\pm2.69 \\
                        \hline \hline
\end{array}
$$
\caption{Parameters appearing in the fit function of the coupling
form factor for $\Lambda_bNB$ vertex.} \label{fitparam}
\renewcommand{\arraystretch}{1}
\addtolength{\arraycolsep}{-1.0pt}
\end{table}
\begin{table}[h]
\renewcommand{\arraystretch}{1.5}
\addtolength{\arraycolsep}{3pt}
$$
\begin{array}{|c|c|c|c|}
\hline \hline
       \mbox{structure}    & c_1 & c_2 (\mbox{GeV$^2$})&c_3     \\
\hline
\mbox{$\gamma_5$} &-0.08\pm0.02&-17.74\pm4.79&0.97\pm0.29 \\
  \hline
  \mbox{$\!\not\!{p}\gamma_5$} &-9.01\pm2.70&-328.82\pm98.65&14.81\pm4.00 \\
  \hline
   \mbox{$\!\not\!{q}\gamma_5$} &-20.04\pm5.81&-1221.76\pm366.53&27.32\pm8.20 \\
  \hline
  \mbox{$\!\not\!{q}\!\not\!{p}\gamma_5$} &0.86\pm0.26&16.63\pm4.82&4.05\pm1.22 \\
                        \hline \hline
\end{array}
$$
\caption{Parameters appearing in the fit function of the coupling
form factor for $\Lambda_cND$ vertex.} \label{fitparam1}
\renewcommand{\arraystretch}{1}
\addtolength{\arraycolsep}{-1.0pt}
\end{table}
\begin{table}[h]
\renewcommand{\arraystretch}{1.5}
\addtolength{\arraycolsep}{3pt}
$$
\begin{array}{|c|c||c|c|}
\hline \hline
     \mbox{structure}     & g_{\Lambda_bNB}(Q^2=-m_{B}^2)&  g_{\Lambda_cND} (Q^2=-m_{D}^2)   \\
\hline
  \mbox{$\gamma_5$} &8.97\pm2.69&0.91\pm0.27 \\
  \hline
  \mbox{$\!\not\!{p}\gamma_5$} &12.31\pm3.57&5.90\pm1.77 \\
   \hline
  \mbox{$\!\not\!{q}\gamma_5$} &15.57\pm4.67&7.34\pm2.20 \\
  \hline
  \mbox{$\!\not\!{q}\!\not\!{p}\gamma_5$} &13.81\pm4.14&5.11\pm1.53 \\
\hline
  \mbox{average } &12.67\pm3.76&4.82\pm 1.44\\
                         \hline \hline
\end{array}
$$
\caption{Values of the $g_{\Lambda_bNB}$  and $g_{\Lambda_cND}$ coupling
constants for different structures.} \label{couplingconstant}
\renewcommand{\arraystretch}{1}
\addtolength{\arraycolsep}{-1.0pt}
\end{table}

At this stage, we  compare our result of the
coupling constant $g_{\Lambda_cND}$ obtained at $Q^2=0$ with that of
Ref.~\cite{Navarra}  for the Dirac structure $\!\not\!{q} \gamma_5$.  At $Q^2=0$, we get the result $g_{\Lambda_cND}=7.28 \pm2.18$ for this structure, which is consistent with the prediction of \cite{Navarra}, i.e.,  $g_{\Lambda_cND}=\sqrt{4\pi}(1.9\pm0.6)=6.74\pm2.12$ within the errors.

To summarize, we have calculated the strong  coupling
constants  $g_{\Lambda_bNB}$ and $g_{\Lambda_cND}$ in the framework of the three-point  QCD sum rules. Our results can be used in the bottom and charmed mesons clouds
description of the nucleon which may be used to explain exotic events observed by different experiments. The obtained results can also be used in analysis of the results of  heavy ion collision experiments like $\overline{P}ANDA$ at FAIR.
These results may also be used in exact determinations of the modifications in the masses, decay constants  and other parameters of the $B$ and $D$ mesons in nuclear medium.

\section{Acknowledgment}
This work has been supported in part by the Scientific and Technological
Research Council of Turkey (TUBITAK) under the research project 114F018.

\end{document}